\documentclass[12pt]{article}
\usepackage{amsmath}
\usepackage{graphicx}
\usepackage{color, authblk}

\newcommand{\fig}[1]{Fig.~\ref{#1}}
\newcommand{\eq}[1]{Eq. (\ref{#1})}

\newcommand{\dxs}{\partial_{x}}

\begin{document}
\title{Asymptotic survival probability of a particle in reaction-diffusion process with exclusion in the presence of traps}
\author[1]{Trilocha Bagarti\thanks{E-mail: bagarti@hri.res.in}}
\affil[1]{Harish-Chandra Research Institute, Chhatnag Road, Jhunsi, Allahabad-211019,India.}

\author[2]{Kalyan Kundu}
\affil[2]{Institute of Physics, Sachivalaya Marg, Bhubaneswar-751005,India.}

\maketitle
\begin{abstract}
Reaction-diffusion process with exclusion in the presence of traps has been studied. The asymptotic survival probability for the case of uniformly distributed random traps shows a stretched exponential behavior. We show that additional correction terms appear in the stretched exponent when exclusion is taken into account. Analytically it is shown to be $\sim t^{1/6}$ which is verified by numerical simulations.
\end{abstract}

\section{Introduction}
Reaction-diffusion processes in the presence of quenched disorder (i.e. traps) has been studied extensively for almost a hundred years. Smulochowsky introduced a reaction diffusion system with traps as a model to study the process of coagulation in colloids\cite{smolu1916}. Reaction diffusion process with traps has been used to study a number of phenomena such as exciton trapping in crystals, recombination of electro-hole, soliton-antisoliton pair, reaction activated by catalysts etc\cite{hav}. Quenched disorder can introduce self-segregation in the reaction process which affects the reaction kinetics\cite{jayan,htait}. It can also induce self-organization of reactant species around the traps\cite{wei89,tait,jayan1}. It has been shown by Balagurov and Vaks \cite{balagurov}, Donsker and Vardhan\cite{donsker} and Grasberger and Procaccia\cite{gras} that the mean survival probability in the asymptotic time limit has a stretched exponential behavior $P(t) \sim \exp(-\alpha_d \rho^{2/(d+2)}t^{d/(d+2)})$, where $\alpha_d$ is a constant that depends on the dimensionality of the space $d$. This slowing down behavior arises due to the occurrence of large trap free regions in space. If traps are located at regular interval the survival probability shows an exponentially decreasing behavior\cite{wi}. It has also been shown that diffusion in the presence of random traps cannot be described by effective diffusion equation\cite{bixon}. 

A reaction diffusion process with traps can be described as $A+T \rightarrow T$ with a rate $\kappa$ where $T$ and $A$ denotes trap and the diffusion particle respectively. For the case of perfect traps we shall consider the limit $\kappa \rightarrow \infty$ for which every particle arriving at the trap vanishes with probability $1$. The case where $\kappa$ is finite is called trapping reaction with partial traps\cite{thm,wi}. In one dimension the mean survival probability in the asymptotic time limit is given by 
\begin{equation}
\begin{aligned}
P(t)= 16\pi^{-3/2}\exp \left(\alpha (Dt)^{1/3}\right)\left[ b \sqrt{Dt} + c(Dt)^{1/6}+O((Dt)^{-1/6})\right]
\label{surv_prob}
\end{aligned}
\end{equation}
where $\alpha= -3\rho^{2/3} \pi^{2/3}2^{-2/3}$, $b=2 \pi \rho 3^{-1/2}$ and $c=17 \pi^{1/3}\rho^{1/3}2^{-1/3}3^{-1/2}9^{-1}$. Here $D$ is the diffusion constant of the reacting particles and $\rho$ is the average number of traps per unit length. Exact expression for $P(t)$ can be obtained in terms of Meijer G functions. This slowing down behavior is observed due to the existence of large trap free regions with small but nonzero probabilities.

In previous studies mutual interaction such as exclusion between particles has not been considered. In this article, we shall investigate the effect of exclusion on the asymptotic survival probability. In a reaction diffusion system the reactants are usually assumed to be point particles. This assumption is valid when volume exclusion it not important i.e. in diffusion processes where the concentration low. However if we consider volume exclusion in the reaction diffusion process with traps, the stretched exponential law gets modified. In our investigation we have found that an additional correction term $\sim t^{1/6}$ appear in the exponent. This is also verified by stochastic simulations. 

Let us denote by $u(x,t)$ the density of the reactant at position $x$ at time $t$. Due to the repulsive nature of exclusion smoothing of density is favored. Reactants particle moving towards a local maxima in the density will experience a local drift away from the local maxima. This repulsive force which arises due to exclusion can be assumed to be proportional to the gradient $\dxs u(x,t)$. We introduce this as a small perturbation in the reaction diffusion equation. The reaction diffusion equation can be written as
\begin{equation}
\partial_t u = \dxs (D \dxs u + \epsilon(\dxs u) u))+\kappa \sum_{i}\delta(x-x_i)u,
\label{diffeqn2}
\end{equation}
where $x_1,x_2\ldots$ denotes the positions of the traps which are uniformly distributed with mean number of traps per unit length $\rho$. The boundary condition is $\partial_x u =0$ at infinity. The parameter $\epsilon$ is a positive constant that determines the strength of exclusion. We note here that exclusion in \eq{diffeqn2} has been introduced in a meanfield way.

\section{Survival probability}
For the case of perfect traps i.e. limit $\kappa \rightarrow \infty$ the problem reduces to solving the diffusion equation \eq{diffeqn2} in a finite domain $[-a, a]$ with absorbing boundary conditions. We have 
\begin{equation}
\partial_t u = \dxs((1+\epsilon u) \dxs u),~~x \in [-a, a], t>0,
\label{diff_eqn}
\end{equation}
where diffusion constant $D$ is assumed unity. The initial concentration $u(x,0)=1$ and the absorbing boundary condition is $u(\pm a, t)=0 \mbox{~for all~} t>0$. The exponent $~t^{1/3}$ in \eq{surv_prob} has been calculated by taking into account only the lowest mode of the eigenvalue equation $D \dxs^2 \psi + \lambda \psi =0$. In the presence of exclusion we have the following nonlinear eigenvalue problem
\begin{equation}
\dxs((1+\epsilon \psi) \dxs \psi)+\lambda \psi =0,~~x \in [-a, a].
\label{egv_eqn}
\end{equation}
To understand the behavior of the survival probability for we need to know how the eigenvalue $\lambda$ get modified for nonzero $\epsilon$. For small $\epsilon$ we can write
\begin{eqnarray}
\psi  = \psi_0 + \epsilon \psi_1 + \epsilon^2 \psi_2 + \ldots, \nonumber \\
\lambda  = \lambda_0 + \epsilon \lambda_1 + \epsilon^2 \lambda_2 + \ldots.
\label{pert_ser}
\end{eqnarray}
Substituting \eq{pert_ser} in \eq{egv_eqn} we obtain
\begin{subequations}
\begin{align}
\label{eq1}
O(1)&:~~\dxs^2\psi_0+\lambda_0 \psi_0 =0,\\
\label{eq2}
O(\epsilon)&:~~\dxs^2\psi_1+\lambda_0 \psi_1 =-\lambda_1 \psi_0-\dxs(\psi_0 \dxs \psi_0).
%\label{eq3}
%O(\epsilon^2)&:~~\dxs^2\psi_2+\lambda_0 \psi_2 =-\lambda_1 \psi_1-\lambda_2\psi_0 -\dxs(\psi_1\dxs \psi_0+\psi_0 \dxs \psi_1).
\end{align}
\end{subequations}
The above set of linear equation can be solved with the absorbing boundary conditions $\psi_m(\pm a)=0,~~m=0,1,2,\ldots$. The eigenvalues and the eigenfunctions of \eq{eq1}are given by
\begin{eqnarray}
\label{sol_pert0}
\psi_0^{(n)}(x) &=& \frac{1}{\sqrt{a}} \left \{ \begin{array}{ll}
                                                              \cos(n\pi x/2a), & \mbox{if $n$ is odd}, \\
                                                              \sin(n \pi x/2a),  & \mbox{if $n$ is even},
                                                           \end{array} \right. \\
\label{sol_pert1}
\lambda_0^{(n)} &=& \frac{n^2 \pi^2}{4a^2}~~\mbox{for}~ n=1,2,\ldots
\end{eqnarray}

Similarly, from \eq{eq2} after setting $\lambda_0 = \lambda_0^{(n)}$ and $\psi_0 = \psi_0^{(n)}$ we have
\begin{equation}
\dxs^2 \psi_1 + \lambda_0^{(n)} \psi_1 = -\lambda_1 \psi_0^{(n)} - \dxs(\psi_0^{(n)} \dxs \psi_0^{(n)}).
\label{eq_ord1}
\end{equation}
Let $\psi_1$ be written as the sum $\psi_1 = \sum_m A_m \psi_0^{(m)}$ which satisfies the boundary condition. From \eq{eq_ord1} we obtain
\begin{equation}
A_m(\lambda_0^{(n)}-\lambda_0^{(m)}) = -\lambda_1 \delta_{m,n}-\int_{-a}^{+a}\dxs(\psi_0^{(n)} \dxs \psi_0^{(n)})\psi_0^{(m)} dx
\label{pertb_A}
\end{equation}
Using \eq{sol_pert0}, \eq{sol_pert1} and \eq{pertb_A} we can write for odd $n$
\begin{equation}
\int_{-a}^{+a}\dxs(\psi_0^{(n)} \dxs \psi_0^{(n)})\psi_0^{(m)} dx = -\frac{\lambda_0^{(n)}}{\sqrt{a}}\mu_{n,m}
\label{expr_mu}
\end{equation}
where $\mu_{n,m} = [4m \cos(n \pi) \sin (m \pi/2)-8n \cos(m \pi/2) \sin (n\pi)]/\pi (m^2-4n^2)$ for $m$ odd otherwise $\mu_{n,m} = 0$. Similarly for even $n$ we can calculate $\mu_{n,m}$. Here we will not require to calculate beyond $\mu_{1,1}$ since we only need $\lambda^{(1)} = \lambda_0^{(1)}+ \epsilon \lambda_1^{(1)}$ to determine the behavior of the asymptotic survival probability. The eigenfunctions and the eigenvalues for $n=1$ can be written as
\begin{eqnarray}
\psi^{(1)} &=& \psi_0^{(1)} +\frac{\epsilon}{\sqrt{a}} \sum_{m \neq 1}^{\infty} \mu_{1,m} \frac{\lambda_0^{(1)}}{\lambda_0^{(1)}-\lambda_0^{(m)}} \psi_0^{(m)}, \nonumber \\
\lambda^{(1)} &=& \lambda_0^{(1)} \left(1+\frac{\epsilon \mu_{1,1}}{\sqrt{a}} \right).
\end{eqnarray}
The greatest contribution to the stretched exponential behavior comes from $n=1$ case. The survival probability can therefore be written as
%from (cf. \eq{sol_surv} and \eq{sol_fin})
\begin{equation}
P_{\epsilon}(t) \sim \int_{0}^{\infty} \exp(-\lambda^{(1)}(x)t) \rho^2 4x \mbox{e}^{-2 \rho x}dx
\label{prob_surv_eps}
\end{equation}
We will use the Laplace method to evaluate \eq{prob_surv_eps} as done by Balagurov and Vaks \cite{balagurov}. The exponent in \eq{prob_surv_eps} can be explicitly written as
\begin{equation}
f(x) = -\frac{\pi^2}{x^2}\left(1+ \frac{\sqrt{2}\epsilon \mu_{1,1}}{\sqrt{x}} \right)t-\rho x.
\end{equation}
The function $f(x)$ attains maximum at 
\begin{equation}
\tilde{x} \simeq x^{*}+\epsilon \frac{5 \mu_{1,1}}{6 \sqrt{2}} \sqrt{x^{*}}
\end{equation}
 where $x^{*} = (2 \pi^2 t/\rho)^{1/3}$. The integral \eq{prob_surv_eps} becomes
\begin{equation}
\begin{aligned}
P_{\epsilon}(t) &\sim  \alpha \exp(f(\tilde{x})),\\
&= \alpha \exp \left(\frac{-3(\rho^2 \pi^2 t)^{1/3}}{2^{2/3}} - \epsilon \mu_{1,1} \left( \frac{\pi}{2} \right)^{1/3} \rho^{5/6} t^{1/6} \right),
\end{aligned}
\label{prob_excl}
\end{equation}
where $\alpha=\left(\frac{\pi}{2 f''(\tilde{x})}\right)^{1/2}\tilde{x},~\mu_{1,1} = 4/3\pi$. 
\section{Numerical simulation and results}
Numerical solution of the diffusion equation \eq{diff_eqn} is obtained by the Langevin simulation of interacting particle system. In our simulation we cannot introduce exclusion between pair of particles by a repulsive interaction. As this is not a lattice simulation where exclusion is incorporated by fixing the upper bound of the site occupancy. The difficult arises here since we have a mean-field description of repulsive interaction in \eq{diff_eqn}. Therefore we need to solve the problem in a self-consistent manner. We consider $N$ particles in $[-a,a]$ with a uniform density $u(x,t)=1$ at time $t=0$. A corresponding self-consistent Langevin equation can be written as
\begin{equation} 
X^{(i)}_{t+dt} = X^{(i)}_t + F(X^{(i)}_t)dt + \sqrt{2Ddt} \xi^{(i)}_t, i=1,2,\ldots N
\label{langv}
\end{equation}
with the force term $F(X^{(i)}_t) = -\epsilon \dxs u(x,t)|_{x=X^{(i)}_t}$ and initial conditions $X^{(i)}_0=\zeta_i$ is a uniform random variable in $[-a,a]$ such that the initial density $u(x,0)$ is constant. The diffusion constant $D=1$ and $\xi^{(i)}_t \sim N(0,1)$ is a Gaussian random variable. 

\begin{figure}
\begin{center}
\includegraphics[angle=-90, width=0.8\textwidth]{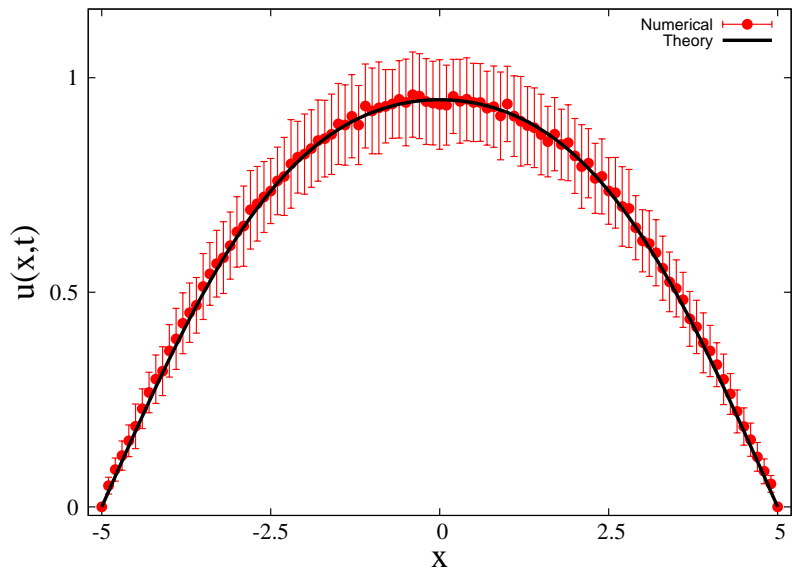}
\caption{Solution obtained from numerical simulation is compared with the theoretical solution $u(x,t)$. Parameter used are $D=1$, $\epsilon=0$ and $t=2.5$.}
\label{fig:sol}
\end{center}
\end{figure}

Evolution of the $N$ particle system is performed self consistently in the following way. In the interval $[-a,a]$ $N$ particles are uniformly distributed initially. Let $x_1,\ldots, x_M$ denote a partition of the interval $[-a,a]$. The density $u(x,0)$ is calculated at $x_i$ for  all $i=0,1,\ldots,M$. The system is evolved using \eq{langv} for time step $dt$ and the density is updated simultaneously. The force term $F(X^{(i)}_t)$ is approximated by using central difference for the gradient$\dxs u(x,t)|_{x=X^{(i)}_t} \simeq (u(x_{n+1},t)-u(x_{n-1},t))/2dx$ where $X^{(i)}_t \in [x_n-dx/2,x_n+dx/2]$. The absorbing boundary conditions $u(\pm a,t)=0$ is incorporated by eliminating each particle that exits the interval $[-a,a]$. The system is evolved for a number of times from the same initial condition. The mean density obtained should satisfies the diffusion equation \eq{diff_eqn} and is the required solution. In \fig{fig:sol} we compare the solution obtained from the numerical simulation for $\epsilon=0$ with the exact solution and find that they are in excellent agreement. 

\begin{figure}
\begin{center}
\includegraphics[angle=-90, width=0.8\textwidth]{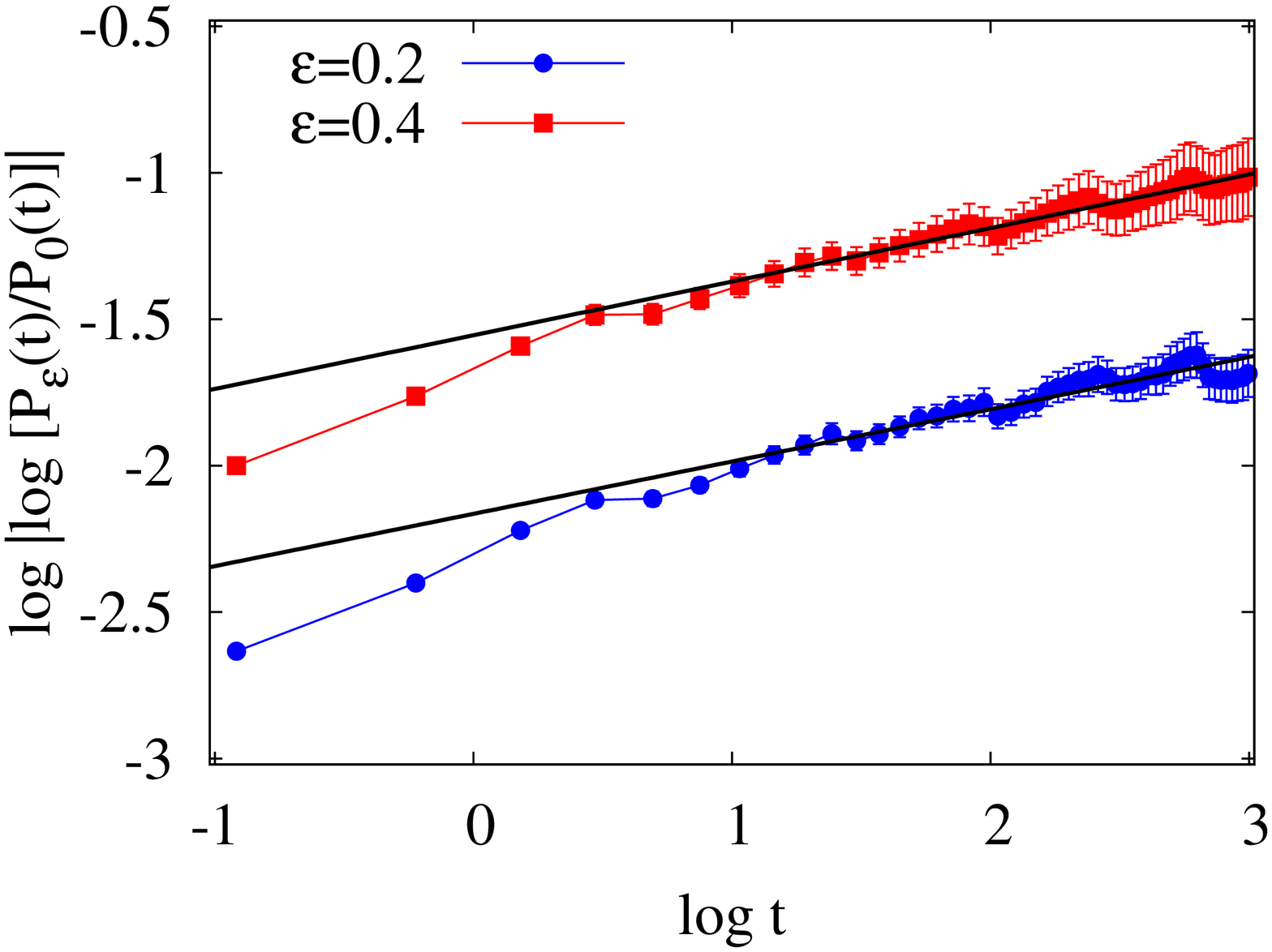}
\caption{A comparison of theoretically predicted slope $1/6$ with the simulation results. The black solid line are least-square fit for $\epsilon = 0.2, 0.4$, $D=1$ and $\rho=1$. The error bars correspond to one standard deviation.}
\label{fig:slope}
\end{center}
\end{figure}

From the analytical expression in \eq{prob_excl} we have
\begin{equation}
\log \left| \log \frac{P_\epsilon(t)}{P_0(t)} \right| = \frac{1}{6} \log t + C,
\label{slope_eq}
\end{equation}
where $C$ is constant. Note that this is true only for small values of $\epsilon$ i.e. $0<\epsilon \ll 1$. The results from the simulation is shown in \fig{fig:slope}. Here we have used the numerical solution to calculate the survival probability $P_\epsilon(t)$. We have taken values $\epsilon = 0.2$ and $0.4$, the diffusion coefficient $D=1$ and the mean number of traps per unit length $\rho$ is unity. Result from our simulation were fitted with straight lines by least square fitting for $\log t \geq 1$. In \fig{fig:slope} the slopes of the straight lines are $0.1784\pm0.0070$ and $0.1831\pm0.0047$ for $\epsilon = 0.2$ and $0.4$ respectively. Although values obtained are slightly higher than the predicted value $0.166\ldots$, the results are satisfactory. We note that the slope $1/6$ is obtained by first order perturbation in $\epsilon$ and it is difficult to obtain an analytical expression with higher order perturbation. For higher values of $\epsilon$ higher order terms will become important in the stretched exponent. However we note that the term $\epsilon \partial_x u$ in \eq{diffeqn2} arising due to exclusion effect is valid for small $\epsilon$.

\section{Conclusion}
We investigated the trapping reaction problem in the presence of exclusion where we found that the survival probability gets modified. We obtained analytically, additional correction term in the stretched exponential law for the survival probability. Our analytical prediction is verified by numerical simulations. Self-consistent Langevin simulation is used to investigate nonlinear reaction diffusion equations. The algorithm used can be further explored to investigate trapping problems in higher dimensions and other (in)homogeneous reaction-diffusion systems with multiple species of reactants.

\newpage

\begin{thebibliography}{13}
\bibitem{smolu1916}M.~V. Smoluchowski, Phys. Z. {\bf 17}, 557 (1916).
\bibitem{hav}S.~Havlin and D.~ben Avraham, Adv. Phys. {\bf 36}, 695 (1987).
\bibitem{jayan}P.~K.~Datta and A. M. Jayannavar, Pramana-J. Phys. {\bf 38}, 257(1992).
\bibitem{htait}H.~Taitelbaum, Physica A {\bf 200}, 155(1993).
\bibitem{wei89}G.~H. Weiss, R.~Kopelman, and S.~Havlin, Phys. Rev. A {\bf 39}, 446 (1989).
\bibitem{tait}H.~Taitelbaum, R.~Kopelman, G.~H. Weiss, and S.~Havlin, Phys. Rev. A {\bf 41}, 3116 (1990).
\bibitem{jayan1}P.~K.~Datta and A.~M.~Jayannavar, Physica A {\bf 184}, 135 (1992).
\bibitem{balagurov}B.~Y. Balagurov and V.~G. Vaks, Sov. Phys. JETP {\bf 38}, 968 (1974).
\bibitem{donsker}M.~D. Donsker and S.~R.~S. Varadhan, Commun. Pure Appl. Math. {\bf 32}, 721 (1979).
\bibitem{gras}P.~Grassberger and I.~Procaccia, J. Chem. Phys. {\bf 77}, 6281 (1982).
\bibitem{thm}T.~M. Nieuwenhuizen and H.~Brand, J. Stat. Phys. {\bf 59}, 53 (1990).
\bibitem{wi}G.~Abramson and H.~S. Wio, Chaos Soliton and Fract. {\bf 6}, 1 (1995).
\bibitem{bixon}M.~Bixon and R.~Zwanzig, J. Chem. Phys. {\bf 75}, 2345 (1981).
\end{thebibliography}
\end{document}